\documentclass[conference]{IEEEtran}

\usepackage{stfloats}
\usepackage{color}
\usepackage{graphicx,color,psfrag}
\usepackage{amsmath}
\usepackage{epsfig}

\newcommand{\qed}{\hspace*{\fill} \fbox{} \par \vspace{\baselineskip}}
\newtheorem{theorem}{Theorem}
\newtheorem{definition}{Definition}
\newcommand{\be}{\begin{equation}}
\newcommand{\ee}{\end{equation}}
\newcommand{\isdef}{\stackrel{\mathrm{def}}{=}}

\IEEEoverridecommandlockouts

\begin{document}
\title{EXIT Chart Approximations using the Role Model Approach}
\author{Jossy Sayir \\ {\tt j.sayir@ieee.org} \\
 Department of Engineering, University of Cambridge, Cambridge CB2 1PZ, UK
\thanks{Part of the material presented in this paper was presented 
at the International Symposium on Communication Theory and Applications
(ISCTA2009) in Ambleside, UK.}}
\maketitle

\begin{abstract}
Extrinsic Information Transfer (EXIT) functions can be measured
by statistical methods if the message alphabet size is moderate
or if messages are true a-posteriori distributions. We propose
an approximation we call {\em mixed information} that constitutes
a lower bound for the true EXIT function and can be estimated
by statistical methods even when the message alphabet
is large and histogram-based approaches are impractical, 
or when messages are not true probability distributions and time-averaging
approaches are not applicable.
We illustrate this with the hypothetical example of a rank-only
message passing decoder for which it is difficult to compute or
measure EXIT functions in the conventional way. We show that the
role model approach \cite{sayir2008} can be used to optimize post-processing for
the decoder and that it coincides with Monte Carlo integration in the
non-parametric case. It is guaranteed to tend towards the optimal
Bayesian post-processing estimator and can be applied in a blind
setup with unknown code-symbols to optimize the check-node operation
for non-binary Low-Density Parity-Check (LDPC) decoders.
\end{abstract}

\section{Introduction}

Extrinsic Information Transfer (EXIT) charts \cite{exit} are a well
known tool to analyze the convergence of iterative decoders and
receivers. While in some cases they can be computed analytically,
one often has to resort to statistical estimation to 
measure the mutual informations plotted in these charts. 
There are two statistical approaches for measuring mutual
information: the histogram-based approach \cite{tenbrink} and the
time-averaging approach \cite{land2004,kliewer2006}. While the
former approach works in all cases but is highly impractical
when the message space is large, the latter approach can only
be applied when the messages correspond to true extrinsic probabilty
distributions over the code alphabet. 

Consider the following example. The sum-product algorithm for
non-binary Low-Density Parity-Check (LDPC) codes
over GF($q$) works by passing probability distributions over
GF($q$) along the edges of a factor graph.
We would like to design a simplified decoder where 
messages are ranked lists of symbols from GF($q$) rather than
probability distributions, a sort of ``Gallager A''
\cite{gallager-thesis} for non-binary codes. Ideally, we would design check and
variable node operations in such a way that the order of decreasing
probabilities in the sum-product messages is retained in the 
ranked list messages of our simplified algorithm.
We are not actually able to design such an algorithm, but we
can analyze the performance achievable by such a
hypothetical algorithm using EXIT charts. In other words,
we are trying to figure out how much information in a ranked list
of probabilities is contained in the rank, or how much information we stand to
loose if we throw away the probability values and retain only the
ranks. In principle, given a measured mutual information, there
exists an algorithm that can exploit it and achieve the performance
we predict in theory, but we concede that it may be impractical
to implement such an algorithm and we currently have no systematic 
way of designing it. Besides, the problem of knowing how much
information is contained in the rank of an a-posteriori 
probability distribution is interesting in its own right.

Let $X$ be a transmitted code symbol from GF($q$), $Y$ be
the corresponding message in the sum-product algorithm, and
$Z$ be the corresponding message in our hypothetical simplified
algorithm, retaining only the orders in the ranked values of $Y$.
For the EXIT
chart, we need to determine the mutual information $I(X;Z)$.
The value of this mutual information will of course depend on the 
exact conditional density $p_{Y|X}$ and 
it is not realistic to compute it analytically. We can however
design a simple experimental setup to measure the mutual 
informations required for the EXIT charts via statistical
methods. However, for $q=16$ there are $16!$ possible orderings
of the code alphabet. Using the histogram method would require
us to measure $16\times 16!$ frequencies in order to estimate the
corresponding conditional probabilties $P_{Z|X}$, which is clearly
not practical. The time-averaging method, on the other hand,
works by computing the expectation of the entropy $H(P_{X|Z=z})$
over all realizations $z$ of the message $Z$. By repeating
this experiment independently in time, ergodicity allows us to
replace the expectation by an average over time. This supposes
however that we know how to compute the a-posteriori distribution
$P_{X|Z=z}$ for every observed message $z$. In our example, computing
the a-posteriori probability of $X$ given a specific probability
ranking $z$ is a difficult problem for which we cannot think of
a computable closed-form solution.

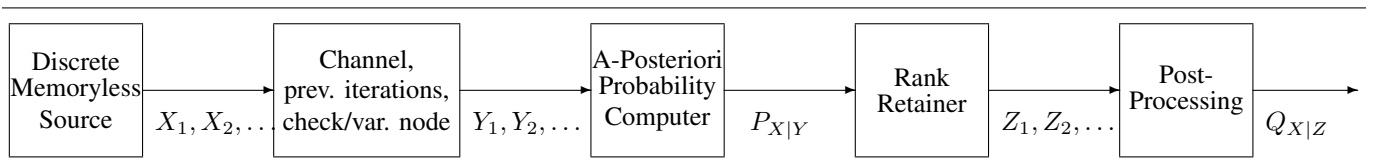
\begin{figure*}[b!]
\vspace*{4pt}
\hrulefill
\begin{center}
\begin{picture}(510,50)
\put(0,0){\framebox(50,50){\shortstack{Discrete\\Memoryless\\Source}}}
\put(50,25){\vector(1,0){50}}
\put(100,0){\framebox(70,50){\shortstack{Channel,\\prev.~iterations,\\check/var.~node}}}
\put(170,25){\vector(1,0){50}}
\put(220,0){\framebox(50,50){\shortstack{A-Posteriori\\Probability\\Computer}}}
\put(270,25){\vector(1,0){50}}
\put(320,0){\framebox(50,50){\shortstack{Rank\\Retainer}}}
\put(370,25){\vector(1,0){50}}
\put(420,0){\framebox(50,50){\shortstack{Post-\\Processing}}}
\put(470,25){\vector(1,0){40}}
\put(55,10){$X_1,X_2,\ldots$}
\put(175,10){$Y_1,Y_2,\ldots$}
\put(375,10){$Z_1,Z_2,\ldots$}
\put(475,10){$Q_{X|Z}$}
\put(280,10){$P_{X|Y}$}
\end{picture}
\caption{Mutual Information Measurement Setup}
\end{center}
\label{fig:fig1}
\end{figure*}
In this paper, we will propose an approximation of the EXIT
function that provides a lower bound for the true mutual
information $I(X;Z)$. This function is computed using a 
parametric model a-posteriori distribution of $X$ given
$Z$ and the true distribution of $X$ given $Y$. It can be
optimized in an adaptive and blind manner, i.e., without requiring
known code symbols $X$, in such a manner as to refine the model
distribution so that it will approach the true
unknown a-posteriori distribution of $X$ given $Z$. 
Our approach uses the role model framework introduced in
\cite{sayir2008}. Its application to EXIT chart approximation
was first described in \cite{sayir2009}. This paper adds
a fresh perspective to this approach by treating the example
of the hypothetical rank-decoder, and draws parallels
between the role model approach, the EM algorithm,
and Monte Carlo integration.

\section{Measurement Setup}

Figure~\ref{fig:fig1} represents our measurement setup for
mutual information. Symbols $X$
are emitted by a discrete memoryless source over the code
alphabet GF($q$) and transmitted over a ``super-channel'' 
consisting of a communications channel, processing performed
at previous iterations, and the check or variable node operation
for which we wish to draw the EXIT curve. The sum-product algorithm
computes the optimal Bayesian message $P_{X|Y}$ at the output of this
channel. It may seem confusing to see this a-posteriori distribution
among the signals in the block diagram of our measurement setup.
In our context,
 we consider this probability-valued message to
be itself a random variable. If the probability distribution computed
by the node is indeed an a-posteriori probability vector, it
is a sufficient statistic for $Y$.

Following this, we have the rank retainer which discards the
probability values from the sum-product message and retains
only the ranked list $Z$ of symbols in GF($q$), as explained in
the introduction. The rightmost box in the figure is an additional
operator that we will use in the mutual information measurement,
labeled ``post-processing''. Its role is to convert the message
$Z$ into the space of probability distributions over
$X$, but not necessarily into the true a-posteriori distribution $P_{X|Z}$.
This is why we use the letter $Q$ instead of $P$ to denote this
message.

If our aim was to compute $I(X;Y)$, then we could use the 
following approach
\begin{align*}
I(X;Y) & = H(X) - H(X|Y) \\
&= H(X) - \sum_y P_Y(y)H(X|Y=y).
\end{align*}
$H(X)$ is known and the expectation in the rightmost term can
be computed through time-averaging due to the ergodicity of
i.i.d.~random processes. This allows us to compute the entropy
of each message $H(P_{X|Y_i=y_i})$, then averaging it over time,
which is essentially the approach descibed in \cite{land2004,kliewer2006}.

In our scenario, our aim is to compute $I(X;Z)$. Since we do not know
how to compute $P_{X|Z}$ easily, we cannot apply the same trick.
We can attack the problem the other way around by writing
out the mutual information as
\[
I(X;Z) = H(Z)-H(Z|X)
\]
but this involves measuring histograms for every possible
value of $Z$, whose alphabet is factorial in $q$ as laid out
in the introduction.

Note that if our post-processing element computes fake
a-posteriori probability distributions $Q_{X|Z}\neq P_{X|Z}$, we
may be tempted to use the same trick as above with these
fake messages. This however gives totally random results as
shown in \cite{sayir2009}. In particular, we could choose
our post-processor to always return a probability of 1 of
getting the symbol 0, irrespective of its input message,
yielding a maximum mutual information even though the resulting
message is independent of $X$ and therefore in reality $I(X;Q_{X|Z}) = 0$.

\section{Mixed Information}

We define the following quantity
\begin{definition}
The {\em mixed information}\footnote{
Note that this quantity was called ``mismatched information'' in 
\cite{sayir2008}. Meanwhile, we have become aware of the large
body of literature on mismatched decoding, where mismatched
information is a well-known term and is defined in a different
manner, e.g., \cite{merhav94}, so we refrain from using that term}
 is defined as
\be
I'(X;Z) \isdef \sum_x \sum_z P(x,z) \log_2 \left(Q(x|z)/P(x)\right).
\label{eq:mixed-inf}
\ee
\label{def:mimsatched-inf}
\end{definition}
Mixed information can be measured via statistical
measurement using an approach parallel to the time averaging
outlined in the previous section:
\begin{equation}
\begin{array}{l}
I'(X;Z) =  \sum_{x,y,z} P(x,y,z) \log_2 \frac{Q(x|z)}{P(x)} \vspace{.2cm} \\ 
 =  H(X) +  \sum_{z,y} P(yz) \sum_x P(x|y) \log_2 Q(x|z)
\end{array}
\label{eq:mixed-inf-avg-2}
\end{equation}
where, again, the expected value in the rightmost term can be
computed via time averaging due to the ergodicity of the i.i.d.~processes
involved.
Unlike the measurement of the true mutual information,
this statistical measurement can be implemented accurately 
even when $Q_{X|Z}\neq P_{X|Z}$.
It is practical even when $Z$ is defined over a large alphabet
since it does not require the computation of histograms for $Z$.

While it is good to know that we can measure mixed information via time averaging,
a crucial  question is how it relates to the true mutual information. Building on
(\ref{eq:mixed-inf-avg-2}), we can write 
\begin{equation}
\begin{array}{l}
I'(X;Z)  =  H(X) + \\
\sum_{y,z} P(yz) \sum_x P(x|y) \left[\log_2\frac{Q(x|z)}{P(x|y)} + \log_2 P(x|y)\right] \vspace{.2cm} \\
 =  H(X) - H(X|Y) - \\
\sum_y\sum_z P(yz) \sum_x P(x|y) \log_2\frac{P(x|y)}{Q(x|z)} \vspace{.2cm} \\
=  I(X;Y) - E_{P_{YZ}}\left[ D(P_{X|Y}||Q_{X|Z})\right]
\end{array}
\label{eq:divergence-expr}
\end{equation}
where we have used a notation from \cite{coverthomas} and expanded
in \cite{sayir2008}
for the expected divergence.
This allows us to state the following theorem:
\begin{theorem}
\be
I'(X;Z) \leq I(X;Y)
\ee
with equality if and only if $Q_{X|Z}= P_{X|Y}$ for all observations $Y$, i.e., if the decoder
under scrutiny is an optimal Bayesian decoder.
\label{th:mixed-bound-opt}
\end{theorem}
{\em Proof: } the proof follows directly from the non-negativity of information divergence.
\qed
In other words, mixed information is always smaller or equal than the
channel mutual information. This is unlike the ``fake'' mutual information
that we mentioned at the end of the previous section,
which could be higher than the channel mutual information, in violation of
the data processing theorem. 

Theorem~\ref{th:mixed-bound-opt} relates mixed information to 
the mutual information over the channel, but it does not tell us how mixed
information relates to the mutual information $I(X;Z)$, which is the one we are
after. For this purpose, we use the following theorem, introduced in
\cite{sayir2008}:
\begin{theorem}[The ``role model'' theorem] If $X$, $Y$ and $Z$ form a 
Markov chain $X-Y-Z$, then
\begin{equation}
\begin{array}{rcl}
\mbox{E}D(P_{X|Y}||Q_{X|Z}) & = & H(X|Z) - H(X|Y) + \vspace{.2cm} \\
& & \mbox{E}D(P_{X|Z}||Q_{X|Z}).
\end{array}
\label{eq:rm-equality}
\end{equation}
In particular, 
\begin{equation}
\mbox{E}D(P_{X|Y}||Q_{X|Z}) \geq H(X|Z) - H(X|Y)
\label{eq:rm-inequality}
\end{equation}
with equality if and only if $Q_{X|Z=z} = P_{X|Z=z}$
for all $z$ for which $P(z)>0$.
\label{th:rolemodel}
\end{theorem}
A direct consequence of this theorem is that mixed information is maximized
for $Q_{X|Z} = P_{X|Z}$, giving the following result:
\begin{theorem}
\be
I'(X;Z) \leq I(X;Z)
\ee
with equality if and only if the post-processing is optimal,
i.e., $Q_{X|Z} = P_{X|Z}$.
\label{th:max-mutinf}
\end{theorem}
The theorem shows that mixed information is a tight lower bound
for the mutual information $I(X;Z)$ and that mixed information
is maximized by the optimal post-processing function.

\section{Role Model Estimation}

The combination of theorems \ref{th:rolemodel} and \ref{th:max-mutinf}
give us a recipe for computing a lower bound approximation to the
EXIT curve in cases where histogram-based approaches are impractical.
In our simplified ranked-list decoder example, all we need to do
is to devise a heuristic probabilistic model $Q_{X|Z}$ of $X$ given
the ranked lists $Z$. Once we establish this model and compute its
corresponding mixed information, Theorem \ref{th:rolemodel} gives
us a method to refine the model and increase the mixed information,
whereby the mixed information tends towards the mutual information 
$I(X;Z)$ and the model tends towards the true a-posteriori probability
distribution. The optimization problem to be solved is a simple
divergence minimization, which is convex in the full set of parameters
$Q_{X|Z}$.

There are some parallels and differences to be drawn between this approach
and Expectation-Maximization (EM) algorithms \cite{baum70,dempster77}.
Both have in common that they design an estimator based on incomplete
observations, in our case $Z$, by bringing the problem back to an
estimation problem based on superior or complete observations, in our
case $Y$, where a Markov condition $X-Y-Z$ ensures the superiority
of $Y$ over $Z$ with respect to estimating $X$. The difference however
is that the EM algorithm ``fabricates'' the complete data and
uses an iterative process to produce an estimator based on the
incomplete data where the complete data has been factored out. 
In contrast, we assume in our setting that the superior data
is available for training purposes in order to train and refine
our estimator based on the inferior data. By doing this, we 
automatically inherit the statistical model for $X$ available
for observations $Y$ and any prior on $X$ that results from this
model. Consequently, the role model approach results in a simple
divergence minimization, mirroring the E step in the EM algorithm
rather than the M step where a divergence is maximized. Note also that
 the divergence maximized in the M step of the EM algorithm
is from the model to the true probability rather than
vice versa as is our case.

Let us now assume that we can adapt the full set of parameters
of the model family of probability distributions $Q_{X|Z}$
and see what the role model approach gives. If we set up 
the Karush-Kuhn-Tucker (KKT) conditions for the divergence minimization
problem, we obtain that the role model estimator boils down
to an evaluation by Monte Carlo integration of the sum
\begin{equation}
P(x|z) = \sum_y P(x|y)P(y|z).
\label{eq:montecarlo}
\end{equation}
The unknown in this equation is $P(y|z)$, which is why
Monte Carlo integration is necessary and
constitutes the best we can hope
to achieve in terms of estimating $P(x|z)$ by observing realizations
of $Y$ and $Z$.

\section{EXIT Charts}

\begin{figure}[h]
\centering
\includegraphics[viewport=81   227   529   564,scale=0.5]{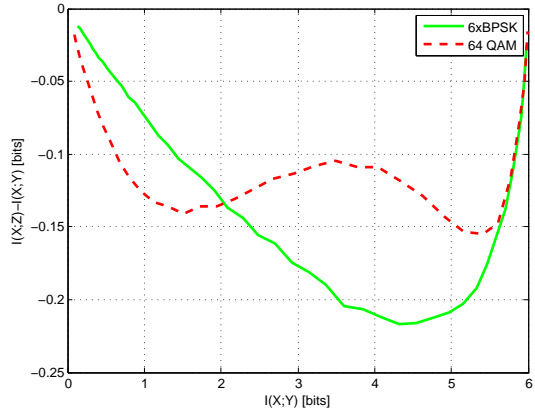}
\caption{Mutual information loss after rank retainer for bitwise
BPSK and for 64-QAM for symbols $X$ over GF(64)}
\label{fig:mutinf_loss}
\end{figure}

As a first step, we were interested in the problem
of finding out how much is lost by retaining the rank of the
a-posteriori probability message in a simple channel. For this,
we are using exactly the setup described in Figure~\ref{fig:fig1}
with a simple communication channel used as the ``super-channel''.
We are generated symbols from GF(64) uniformly at random and
transmitted them over an Addititive White Gaussian Noise (AWGN)
channel, once modulated as six BPSK symbols per source symbol
using natural mapping,
and once modulated using 64-QAM. The loss of mutual information
measured is plotted in Figure~\ref{fig:mutinf_loss}.
The measurement shows that surprisingly little information is
contained in the actual values of the probabilities in the
a-posteriori distributions computed, whereas most of the information
is in the ranked list of symbols. The loss of mutual information
when retaining only the ranked list is less than $1/4$ bit in all
cases. It is generally smaller for 64-QAM than for bitwise BPSK,
except at very low SNR where the loss is higher for 64-QAM.

For the EXIT chart measurement for variable nodes, we used
the 64-QAM modulated AWGN channel with a signal to noise ratio
of $E_s/N0 = 16.3 dB$ ($E_b/N0 = 8.5 dB$). All our EXIT curves
are for check node degree $d_c=4$ and variable node degree $d_v=2$
regular LDPC codes over GF(64). The quality of the
EXIT analysis depends on the availability of a realistic parametric
model for the distributions of messages coming from previous
iterations and there is much literature covering this issue,
e.g., \cite{bb04,bb06}. Since we are measuring a hypothetical
algorithm, we opted at this stage to use simple 64-QAM modulated
symbols through an AWGN channel as our incoming messages. We
are aware that this casts some doubt over the validity of any
performance predictions, but any approximation related to our message
distribution is overshadowed by the far greater approximation
related to the fact that we are measuring a hypothetical algorithm
that we cannot actually implement and are currently investigating
a concept rather than aiming to predict exact performance.

For the EXIT chart of the sum-product algorithm, we measure the
mutual information of our generated incoming messages using time
averaging, then apply the variable node rule of the sum-product
algorithm and measure the mutual information of our extrinsic
messages again using time averaging. For the EXIT chart of our
hypothetical algorithm, we apply the rank retainer to the a-posteriori
messages computed based on our generated graph messages. We then
use our role model framework to optimize a post-processing distribution
$Q_{X|Z}$ of the symbols given the ranked lists. This distribution
is used both to compute a lower bound on the EXIT curve and in the
variable node operation, since we do not know any other way of 
implementing a variable node operation at this stage that processes
incoming ranked lists and computes a ranked list. Instead, we 
implement the usual sum-product variable node rule but use the
optimized $Q_{X|Z}$ instead of the original $P_{X|Y}$ as the 
incoming messages. The resulting message computed by the variable node
 is again passed through a rank retainer so as to get a mapping
from lists to lists as intended, and the mutual information of the
final extrinsic list-valued message is measured again using time-averaging
and the role model framework.

\begin{figure}[h]
\centering
\includegraphics[viewport=81   227   529   564,scale=0.5]{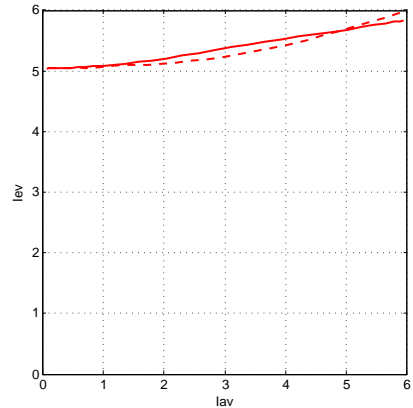}
\caption{EXIT chart for variable nodes $d_v=2$ over GF(64), $E_b/N_0=8.5 dB$, 64-QAM over AWGN, for the sum-product
algorithm (dashed curve) and for the hypothetical rank-based algorithm
(continuous curve)}
\label{fig:exit_var}
\end{figure}

\begin{figure}[h]
\centering
\includegraphics[viewport=81   227   529   564,scale=0.5]{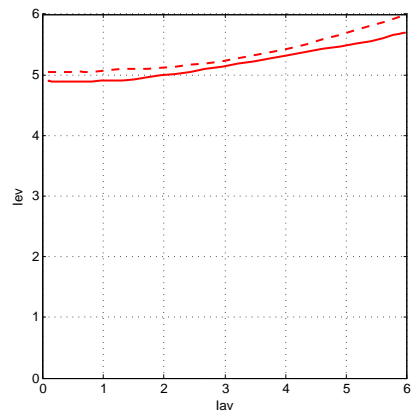}
\caption{EXIT curve as in Figure~\ref{fig:exit_var} but where the 
extrinsic information for the rank-based decoder is plotted in function
of the same mutual information as for the sum-product decoder}
\label{fig:exit_var_corr}
\end{figure}

The EXIT curves measured for variable nodes are plotted in 
Figure~\ref{fig:exit_var}.
Surprisingly, the curve for the rank-based variable node is in part above
the sum-product curve, in apparent breach of the data processing theorem.
This is merely an illusion because the mutual informations on the
x-axis do not correspond to the same incoming messages. The x-axis
for the rank-based algorithm is in effect ``warped'' to its advantage, i.e.,
you need to start off with more informative messages if you want the
resulting ranks to contain as much information as the equivalent
full messages used by the sum-product algorithm. Figure~\ref{fig:exit_var_corr}
shows an equivalent EXIT curve where the extrinsic information for the
rank-based decoder is plotted in function of the mutual information
of the original incoming messages before the rank retainer and shows
that there is no breach of the data processing theorem as we feared.
More interestingly, both figures show that the extrinsic mutual information
for the rank-based decoder does not converge to 6 bits as the a-priori
information goes to 6. This implies that any decoder that uses rank-based
variable nodes is bound to have a decoder-induced error floor. 

\begin{figure}[h]
\centering
\includegraphics[viewport=81   227   529   564,scale=0.5]{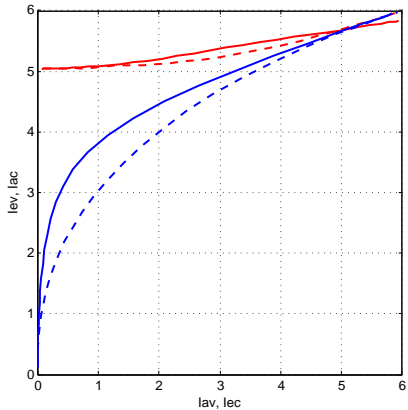}
\caption{Full EXIT chart for regular (4,2) LDPC over GF(64), $E_b/N_0=8.5 dB$, 64-QAM over AWGN, for the sum-product
algorithm (dashed curve) and for the hypothetical rank-based algorithm
(continuous curve)}
\label{fig:exit}
\end{figure}
The resulting full EXIT chart with all measured EXIT functions including
the check node curves for $d_c=4$ is in Figure~\ref{fig:exit}. 
The curve of the rank-based check node shows a significant loss with
respect to the sum-product check node. However, in the important region
where the decoder converges to the error-free case, the curves have the
same slope and become barely distinguishable. Our results are an 
indication that perhaps a hybrid decoder that somehow implements a
rank-based operation in the check nodes while remaining in the probability
domain in the variable nodes may be able to attain good performance.
Since at this point we have neither a method for designing such a 
check node nor a precise setup for predicting performance, this 
indication is to be taken as an encouragement for futher research
rather than an accurate prediction.

\section{Conclusion}
We have introduced {\em mixed information} and shown that it is
a tight lower bound both for the mutual information over the channel
and for the mutual information between the transmitted symbols and
the output of a sub-optimal decoder. Mixed information
can be measured effectively via time averaging.
This allows us to draw a lower bound for the EXIT
function, enabling
the design of codes that are guaranteed to converge and matched to a
practical simplified decoder.

Furthermore, since mixed information is maximized by an optimal
post-processing function, i.e., when the output of the post-processor is
the true a-posteriori probability distribution given the sub-optimal
decoder output, it can serve as a design tool
for the post-processing stage. By choosing the 
parameters of the post-processing function so as to maximize mixed
information, we are guaranteed to approach the optimal post-processor
that gives the best possible performance for a given choice of the
simplified decoder component.

We have illustrated the application of mixed information and
the role model approach by using it to estimate the EXIT function
of a hypothetical check node that retains only the symbol ranks
in an ordered list of probablities in the messages of the sum-product
algorithm. We have shown how for non-parametric estimation the
role-model framework is equivalent to Monte Carlo integration, but
whereas the latter cannot be applied to parametric models the former
can.

\section*{Acknowledgments}
The author is indebted to his office mate Simon Hill for his help in
clarifying the relation between the role model approach and Monte
Carlo integration, and hopes that this acknowledgment in addition
to multiple daily supplies of Turkish coffee will suffice to earn him
the much-coveted title of ``scholar and gentleman'' that Simon 
likes to bestow upon those in his esteem.


\end{document}